\documentclass[aps,prd,preprint,showpacs,showkeys,superscriptaddress]{revtex4}

\usepackage{latexsym}
\usepackage{amsmath,amssymb}
\usepackage[dvips]{graphicx}
\usepackage{subfigure}
\usepackage[dvips]{hyperref} 

\begin{document}


\title{Traversable wormhole in the deformed Ho\v{r}ava-Lifshitz gravity}

\author{Edwin J. Son}
\email[]{eddy@sogang.ac.kr}
\affiliation{Center for Quantum Spacetime, Sogang University, Seoul 121-742, Korea}

\author{Wontae Kim}
\email[]{wtkim@sogang.ac.kr}
\affiliation{Center for Quantum Spacetime, Sogang University, Seoul 121-742, Korea}
\affiliation{Department of Physics, Sogang University, Seoul 121-742, Korea}
\affiliation{School of Physics, Korea Institute for Advanced Study, Seoul 130-722, Korea}

\date{\today}

\begin{abstract}
Asymptotically flat wormhole solutions are found in the deformed Ho\v{r}ava-Lifshitz gravity.
It turns out that higher curvature terms cannot play the role of exotic matters which are crucial to form a traversable wormhole, and external exotic sources are still needed.
In particular, the exotic matter behaves like phantom energy if the Kehagias-Sfetsos vacuum is considered outside the wormhole.
Interestingly, the spherically symmetric setting makes the matter and the higher curvature contribution satisfy four-dimensional conservation of energy in the covariant form.
\end{abstract}



\maketitle

\section{Introduction}

Recently, an ultraviolet (UV) completion of general relativity, motivated by the Lifshitz theory in the condensed matter physics~\cite{lifshitz}, has been proposed by Ho\v{r}ava~\cite{horava}.
The Ho\v{r}ava-Lifshitz (HL) gravity is established as a power-counting renormalizable theory at the cost of the violation of the Lorentz symmetry, which is responsible for the fact that the HL theory is not invariant under the full diffeomorphism group of general relativity but under its subgroup, called the foliation-preserving diffeomorphism.
However, the full diffeomorphism is somehow recovered in the infrared (IR)
limit, though a mechanism for recovering the full diffeomorphism or the renormalization group flow is yet unsolved issue.
The HL gravity has been intensively studied in the area of black hole physics~\cite{lmp,cco,cy,ks,mipark,ysmyung,mann,gh,ir,tc,kk:bh,majhi} and cosmology~\cite{ts,calcagni,kk,mukohyama,brandenberger,cpt,ww,bps,ls,sj,sk,sk2}.

On the other hand, a spacetime wormhole is a widely known object providing a conceivable method for rapid interstellar travel.
In 1988, Morris and Thorne studied traversable wormholes in a realistic manner~\cite{mt}.
They pointed out some problems of Schwarzschild wormholes to be used for interstellar travel and listed the desired properties that traversable wormholes should possess.
One important property for traversable wormholes is obviously that there should be no horizon, since it would prevent two-way travel, which is actually a critical problem of Schwarzschild wormholes.
Considering matter sources, we can get rid of the horizon out of the geometry, though the matter which is needed to make wormholes traversable violates (some of) the desired energy conditions.
This kind of matter is called the \emph{exotic} matter.

Motivated by the fact that the higher curvature terms in HL cosmology can make negative contributions to energy density~\cite{ts}, we are trying to see if they can also play the role of exotic matters.
In this respect, we will find traversable wormhole solutions in HL gravity, considering the ``detailed balance'' condition (DBC) with IR modification.
The deformed HL gravity is considered because the asymptotic flatness requires the IR modification to HL theory, which was a key to have an asymptotically flat solution known as Kehagias-Sfetsos (KS) black hole~\cite{ks}.
In the deformed HL gravity, however, it turns out that the exotic sources are still needed, and the higher curvature contributions are not exotic at all.

The present paper is organized as follows.
We catch a glimpse of HL gravity and KS vacuum in Sec.~\ref{sec:HL}. 
Then, the energy densities and pressures are obtained for the matter source and the higher curvature contribution in Sec.~\ref{sec:exotic}, where it turns out that the higher curvature terms cannot play the role of exotic source, considering the flaring-out condition.
In Sec.~\ref{sec:traversable}, some conditions for traversable wormholes are found and three types of wormhole solutions are examined.
Finally, in Sec.~\ref{sec:discussion}, some comments and discussion will be given.

\section{A glimpse of Ho\v{r}ava-Lifshitz gravity}
\label{sec:HL}
Considering Arnowitt-Deser-Misner decomposition of the
metric with $ds^2 = - N^2 c^2 dt^2 + g_{ij} (dx^i + N^i dt) (dx^j
+ N^j dt)$~\cite{adm} and an anisotropic scaling between time and space, $t \to b^{z}\, t$ and $x^i \to b\, x^i$, 
the HL gravity of $z=3$ with the \emph{softly}
broken DBC is given by~\cite{horava}
\begin{equation}
\label{action}
\begin{aligned}
I_\text{HL} = \int dt d^3x \sqrt{g} N \bigg[ & \frac{2}{\kappa^2} \left( K_{ij} K^{ij} - \lambda K^2 \right) - \frac{\kappa^2}{2\zeta^4} \left( C_{ij} - \frac{\mu\zeta^2}{2} R_{ij} \right) \left( C^{ij} - \frac{\mu\zeta^2}{2} R^{ij} \right) \\
  &+ \frac{\kappa^2\mu^2}{8(3\lambda-1)} \left( \frac{4\lambda-1}{4} R^2 + ( \omega - \Lambda_W ) R + 3 \Lambda_W^2 \right)\bigg],
\end{aligned}
\end{equation}
where $K_{ij} \equiv \frac{1}{2N} \left[ \dot{g}_{ij} - \nabla_i
N_j - \nabla_j N_i \right]$ is the extrinsic curvature at $t=
\text{constant}$ hypersurface, and the dot denotes the derivative
with respect to time $t$. Here, $g_{ij}$, $R$, and $\nabla_i$ are the
metric, the intrinsic curvature, and the covariant derivative in
the three-dimensional hypersurface, respectively.
In addition, $C_{ij}$ is the Cotton-York tensor defined by
\begin{equation}
C^{ij} = \varepsilon^{ik\ell} \nabla_k \left( R_\ell^j - \frac14 \delta_\ell^j R \right),
\end{equation}
$\kappa^2$ is a coupling related to the Newton constant
$G_N$, and $\lambda$ is an additional dimensionless coupling
constant. Note that $\omega$ represents the IR
modification which is essential to have asymptotically flat
solutions. The coupling constants $\mu$, $\Lambda_W$, and $\zeta$
come from the three-dimensional Euclidean topologically massive
gravity action~\cite{djt},
\begin{equation}
W = \mu \int d^3x \sqrt{g} (R-2\Lambda_W) + \frac{1}{\zeta^2} \int \chi(\Gamma),
\end{equation}
where $\chi(\Gamma)$ represents the gravitational Chern-Simons term.
Then, the scaling of couplings can be obtained as $\kappa^2\to \kappa^2$, $\mu\to b^{-1}\mu$, $\Lambda_W\to b^{-2}\Lambda_W$, $\zeta\to\zeta$ and $\omega\to b^{-2}\omega$.
Note that identifying the fundamental constants with
\begin{equation}
\label{fund:const}
c = \frac{\kappa^2}{4} \sqrt{\frac{\mu^2(\omega-\Lambda_W)}{3\lambda-1}}, \qquad G_N = \frac{\kappa^2c^2}{32\pi}, \qquad \Lambda = -\frac{3\Lambda_W^2}{2(\omega-\Lambda_W)},
\end{equation}
the Einstein-Hilbert action can be recovered in the IR limit with $\lambda=1$:
\begin{equation}
\label{act:EH}
\begin{aligned}
I_\text{EH} &= \frac{c^3}{16\pi G_N} \int d^4x \sqrt{-\mathcal{G}} \left[ \mathcal{R} - 2\Lambda \right] \\
  &= \frac{c^2}{16\pi G_N} \int dt d^3x \sqrt{g} N \left[ K_{ij} K^{ij} - K^2 + c^2 \left( R - 2\Lambda \right) \right],
\end{aligned}
\end{equation}
where $\mathcal{G}$ and $\mathcal{R}$ are the metric and curvature scalar of four-dimensional spacetime.
So, we will assume $\lambda=1$ to keep the Einstein limit and $\Lambda_W=0$ to consider an asymptotically flat spacetime.
Then, the HL action~\eqref{action} can be split by $I_\text{HL} = I_\text{EH} + I_\text{HC}$ for the purpose of considering $I_\text{HC}$ as contributions to matter-energy in what follows and eventually investigating the possibility to get a traversable wormhole without any external exotic sources.

Now, varying the total action $I_\text{tot} = I_\text{EH} + I_\text{HC} + I_m$ with a static, spherically symmetric metric ansatz,
\begin{equation}
ds^2 = - e^{2\Phi(r)} c^2 dt^2 + \frac{dr^2}{1-f(r)/r} + r^2 \left( d\theta^2 + \sin^2\theta d\phi^2 \right),
\end{equation}
the equations of motion are obtained as
\begin{align}
& \frac{4}{\kappa^2r^2} e^{-2\Phi} f' = T^{tt}, \label{eom:t} \\
& -\frac{4c^2}{\kappa^2r^2} (1-f/r) \left[ \frac{f}{r} - 2 \left( r - f \right) \Phi' \right] = T^{rr}, \label{eom:r} \\
& -\frac{4c^2}{\kappa^2r^4} \left[ \frac12 r (f/r)' - r \left( r - f \right) \left( \Phi'^2 + \Phi'' \right) - \left( r - f - \frac12 r^2 (f/r)' \right) \Phi' \right] = T^{\theta\theta} = T^{\phi\phi} \sin^2 \theta, \label{eom:theta}
\end{align}
where $T^{\mu\nu} = T_\text{HC}^{\mu\nu} + T_m^{\mu\nu} \equiv -2 \left( \delta I_\text{HC} / \delta \mathcal{G}_{\mu\nu} + \delta I_m / \delta \mathcal{G}_{\mu\nu} \right)$ are the total stress-energy tensors and the prime denotes the derivative with respect to $r$.
Here, the higher curvature contributions of them are derived as
\begin{align}
T_\text{HC}^{tt} = & -\frac{\kappa^2\mu^2}{16c^2r^2} e^{-2\Phi} \left( \frac{f^2}{r^3} \right)', \\
T_\text{HC}^{rr} = & -\frac{\kappa^2\mu^2}{16r^7} \left( r - f \right) \left[ f^2 + 4 r f \left( r - f \right) \Phi' \right], \\
T_\text{HC}^{\theta\theta} = & T_\text{HC}^{\phi\phi} \sin^2 \theta \notag \\
  = & -\frac{\kappa^2\mu^2}{16r^5} \left[ f \left( \frac{f}{r^2} \right)' + \frac{2f}{r} \left( r - f \right) \left( \Phi'^2 + \Phi'' - \Phi' \right) - \left( \frac{f}{r} \right)' (-2r+3f) \Phi' \right].
\end{align}
Note that the higher curvature terms do not contain the parameter $\zeta$, because the above spherically symmetric metric ansatz makes the Cotton-York tensor vanish.
So, in some sense, the static, spherically symmetric metric reduces our model to effectively HL gravity of $z=2$.

\section{Flaring-out condition and exotic matter}
\label{sec:exotic}
\begin{figure}[pbt]
\centering
  \subfigure[]{\label{fig:embed:a}\includegraphics[width=0.45\textwidth]{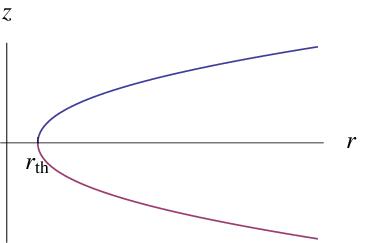}} \qquad
  \subfigure[]{\label{fig:embed:b}\includegraphics[width=0.45\textwidth]{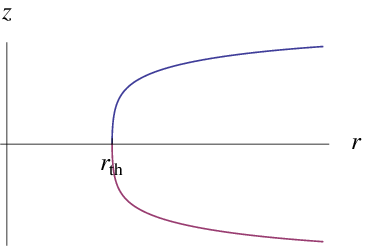}}
  \caption{\label{fig:embed}
  The shape of wormhole is depicted for the case of \subref{fig:embed:a} $f(r) = -r^3\omega + \sqrt{r^3(r^3\omega^2+B)}$, \subref{fig:embed:b} $f(r) = -r^3\omega + \sqrt{r^3(r^3\omega^2+B(r/r_\text{th})^2 e^{-\sqrt{\omega}(r-r_\text{th})})}$.
}
\end{figure}

Since the traversable wormhole is known to be accompanied by the so-called exotic matter in general relativity (GR), the flaring-out condition for the shape of wormholes is introduced to see if the external source and the higher curvature terns are exotic or not.
Following Ref.~\cite{mt}, to study the wormhole geometry, we embed a slice of $t=\text{constant}$ and $\theta=\pi/2$ in an axially symmetric Euclidean space:
\begin{equation}
ds^2 = \frac{dr^2}{1-f/r} + r^2 d\phi^2 = dz^2 + dr^2 + r^2 d\phi^2.
\end{equation}
Then, we have a relation for the embedding function of $dz/dr = \pm \sqrt{f/(r-f)}$, and the wormhole throat is defined as the minimum radius $r_\text{th}$ at which its embedded surface is vertical, \textit{i.e.}  $dz/dr$ diverges as seen in Fig.~\ref{fig:embed}, so that $f(r_\text{th})=r_\text{th}$.
Now, a flaring-out condition $d^2r/dz^2>0$ should be required near the throat:
\begin{equation}
\label{cond:flaring-out}
\frac{d^2r}{dz^2} = -\frac{r^2(f/r)'}{2f^2} > 0 \qquad \text{near } r=r_\text{th},
\end{equation}
so that $(f/r)$ should be a decreasing function at least near the throat and eventually vanish as $r\to\infty$ due to the asymptotic flatness.

Next, to see if the higher curvature terms can play the role of the exotic matter, a dimensionless function is defined as~\cite{mt}
\begin{equation}
\zeta \equiv \frac{\tau-\rho}{|\rho|}.
\end{equation}
If $\zeta>0$, then the energy density is less than the tension so that a certain energy condition should be violated.
In this manner, a matter provided by $\zeta>0$ is called an exotic matter.
Now, the energy density and pressure contributions are obtained from ${T_\text{HC}}^\mu_{~\nu} = \text{diag}(-\rho_\text{HC},-\tau_\text{HC},p_\text{HC},p_\text{HC})$ as
\begin{align}
\rho_\text{HC} = & -\frac{\kappa^2\mu^2}{16r^2} \left( \frac{f^2}{r^3} \right)', \\
\tau_\text{HC} = & \frac{\kappa^2\mu^2}{16r^6} \left[ f^2 + 4 r f \left( r - f \right) \Phi' \right], \\
p_\text{HC} = & \frac{r}{2} \left[ (\rho_\text{HC}-\tau_\text{HC})\Phi' - \tau_\text{HC}' \right] - \tau_\text{HC},
\end{align}
where $\rho_\text{HC}$, $\tau_\text{HC}$, and $p_\text{HC}$ are the energy density, the radial tension, and the lateral pressure, respectively.
It is obvious that the higher curvature contribution to the energy density is positive if $(f^2/r^3)'<0$.
Then, the exotic function for the higher curvature contribution is given by
\begin{equation}
\zeta_\text{HC} = \frac{\tau_\text{HC}-\rho_\text{HC}}{|\rho_\text{HC}|} = -\frac{4f^2}{|3f-2rf'|} \frac{d^2r}{dz^2} + \frac{4r(r-f)}{|3f-2rf'|} \Phi'.
\end{equation}
Note that the finiteness of $\rho_\text{HC}$ yields $(3f-2rf')\ne0$, and if $(1-f/r)\Phi'\to0$ near the throat, then the flaring-out condition reads $\zeta_\text{HC}<0$ near $r=r_\text{th}$, which means the higher curvature terms is not exotic at least near the throat.
Indeed, let $\Phi' = \phi(r)/(1-f/r)^n$ with $\phi(r_\text{th})>0$, then $\Phi$ goes to $-\infty$ as $r$ goes down to $r_\text{th}$ for $n\ge1$, so that the lapse function $e^{\Phi}$ vanishes.
The vanishing lapse function reports that there is a horizon, which forbids two-way travel.
Since we want to find a two-way traversable wormhole, $n$ should be less than one and $(1-f/r)\Phi'\to0$ so that the higher curvature terms cannot play the role of the exotic matter.

Actually, a wormhole could be constructed without any matter source $T_m^{\mu\nu}$, though it is not traversable.
When we remove the matter source, $T_m^{\mu\nu}=0$, then the solution is obtained as
\begin{equation}
f = -r^3\omega + \sqrt{r^3(r^3\omega^2+B)}, \qquad \Phi = C + \frac12 \ln (1-f/r),
\end{equation}
where $B$ and $C$ are constants of integration.
Here, the asymptotic flatness yields $C=0$, and then the solution is exactly the same with the KS solution as expected~\cite{ks}.
Note that the exotic function of the higher curvature terms vanishes in this case, $\zeta_\text{HC}=0$, which is a critical case.
Now, we have a KS wormhole [Fig.~\ref{fig:embed:a}], but it has a horizon at the throat like Schwarzschild wormhole in GR~\cite{mt}, which is why it is not traversable.

To check whether the external source is also exotic or not,
let us now replace the function $f(r)$ by $b(r)$ through the relation
\begin{equation}
\label{f}
f(r) = -r^3\omega + r \sqrt{r^3(r^3\omega^2+b(r))}
\end{equation}
for convenience.
Then, $b(r_\text{th})$ should be positive since $1-f(r_\text{th})=0$, and the energy density and pressure of the external source can be obtained from ${T_{m}}^\mu_{~\nu} = \text{diag}(-\rho,-\tau,p,p)$ as
\begin{align}
\rho = & \frac{\kappa^2\mu^2b'}{16r^2}, \label{rho} \\
\tau = & \frac{\kappa^2\mu^2}{16r^3} \Big[ -4r^3\omega^2 - b + 4r\omega\sqrt{r(r^3\omega^2+b)} \notag \\
  & \qquad - 4 \sqrt{r(r^3\omega^2+b)} \left( 1 + r^2\omega - \sqrt{r(r^3\omega^2+b)} \right) \Phi' \Big], \label{tau} \\
p = & \frac{r}{2} \left[ (\rho-\tau)\Phi' - \tau' \right] - \tau.
\end{align}
Note that the expressions for $\rho$ and $p$ are the same as those in GR up to some factor~\cite{mt}, while $\tau$ has quite different expression, and the energy density is positive when $b'>0$.
That is, the function $b(r)$ should be monotonically increasing if the energy density is assumed to be positive.
Then, the exotic function is given by
\begin{equation}
\zeta = \frac{\tau-\rho}{|\rho|} = \frac{4}{r^4} \left( f + r^3\omega \right) \left[ \frac{f^2}{|b'|} \frac{d^2r}{dz^2} - \frac{r(r-f)}{|b'|} \Phi' \right].
\end{equation}
Note that the flaring-out condition tells us that the external source is exotic near the throat.
In other words, the external exotic sources are still needed to make wormholes traversable in our deformed HL gravity.

\section{Constructing a traversable wormhole}
\label{sec:traversable}
As mentioned in the previous section, a traversable wormhole should not possess a horizon.
To find a condition for it, let us introduce a new radial coordinate $\ell$:
\begin{equation}
d\ell = \pm \frac{dr}{\sqrt{1-f/r}}
\end{equation}
with $\ell=0$ at the throat.
Note that $|\ell(r)|$ reads the proper distance from the throat to $r$, and the throat ($\ell=0$) connects the upper spacetime ($\ell>0$) with the lower spacetime ($\ell<0$).
Then, the line element is written as
\begin{equation}
ds^2 = -e^{2\Phi} c^2 dt^2 + d\ell^2 + r^2(\ell) \left( d\theta^2 + \sin^2 \theta d\phi^2 \right)
\end{equation}
with $r(\ell)\ge r(\ell=0)=r_\text{th}$ by definition.
Now, it is explicitly seen that there should be a horizon when $e^{2\Phi}\to0$ at $r=r_h\ge r_\text{th}$.
In other words, all traversable wormholes should have a finite $\Phi(r)$ for all $r\ge r_\text{th}$.

In addition, we want to find an asymptotically flat wormhole solution; however, if $b(r) \gtrsim r^3\omega^2$ for large $r$, the asymptotic flatness cannot be achieved.
Indeed, if $b(r)\gg r^3\omega^2$, then $f/r \approx 1 - \sqrt{rb(r)} + O(r^2\omega^2)$, and if $b(r) \approx \alpha r^3\omega^2 \left[ 1 + \epsilon(r) \right]$ with a constant $\alpha\ne0$ and a function $\epsilon(r\to\infty)\to0$, then $f/r \approx 1 - (\sqrt{1+\alpha}-1) r^2\omega \left[ 1 + O(\epsilon(r)) \right]$.
In both cases, the function $f/r$ diverges, and we see that the asymptotic flatness is guaranteed only for the case of $b(r) \ll r^3\omega^2$ at large $r$.
Now, we arrive the final condition, $b(r)\ll r\omega$ for large $r$, from the asymptotic flatness, since the function $f/r$ approximates to $f/r \approx b(r)/2r\omega$.
In what follows, three types of asymptotically flat, traversable wormhole solutions are suggested.

\begin{figure}[pbt]
\centering
  \subfigure[\ Exponential case]{\label{fig:exotic:a}\includegraphics[width=0.45\textwidth]{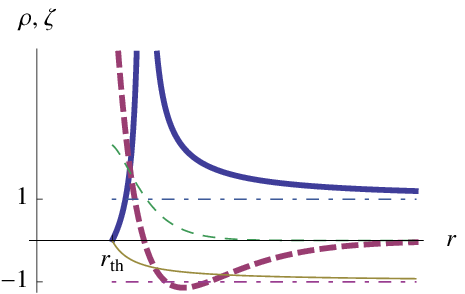}} \qquad
  \subfigure[\ WH + KS solution]{\label{fig:exotic:b}\includegraphics[width=0.45\textwidth]{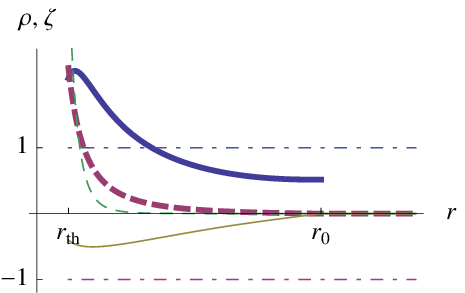}} \\
  \subfigure[\ WH + Minkowski vacuum]{\label{fig:exotic:c}\includegraphics[width=0.45\textwidth]{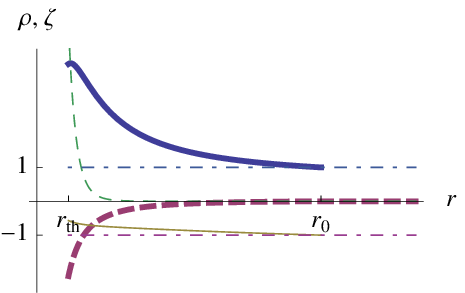}}
  \caption{\label{fig:exotic}
  The exotic functions (solid lines) and the energy densities (dashed lines) of the exotic source (thick lines) and the higher curvature contribution (thin lines) are plotted for the case of \subref{fig:exotic:a} $b(r) = B(r/r_\text{th})^2 e^{-\sqrt{\omega}(r-r_\text{th})}$ and $\Phi(r)=0$, \subref{fig:exotic:b} $b(r) = B \left[ 1 + \sin (\pi/16) \left( r/r_\text{th} - 1 \right) \right]$ and $\Phi(r) = (1/2) \ln (1-f(r_0)/r_0) + (1-r/r_0) (r_0f'(r_0)-f(r_0))/2(r_0-f(r_0))$ for $r\le r_0 \equiv 9r_\text{th}$ and $b(r) = 2B$ and $\Phi = (1/2) \ln (1-f(r)/r)$ for $r>r_0$, and \subref{fig:exotic:c} $b(r) = B \left[ 1 - \sin (\pi/16) \left( r/r_\text{th} - 1 \right) \right]$ for $r\le r_0$ and $b(r) = 0$ for $r>r_0$ and $\Phi(r)=0$.
}
\end{figure}
First, we can consider a solution with an exotic source falling off rapidly as $r$ grows to infinity.
We choose $b(r) = B(r/r_\text{th})^2 e^{-\sqrt{\omega}(r-r_\text{th})}$ and $\Phi(r) = 0$, then we have the energy density of the exotic source as follows:
\begin{equation}
\rho = \frac{\kappa^2\mu^2B(2-\sqrt{\omega}r)}{16 r_\text{th}^2 r} e^{-\sqrt{\omega}(r-r_\text{th})}.
\end{equation}
Note that the energy density of the exotic matter decays exponentially to zero as desired, but it becomes (small) negative for $r>2/\sqrt{\omega}$, though it may be positive near the throat [see Fig.~\ref{fig:exotic:a}].
This kind of wormhole require an exotic matter spreading out all the spaces, which seems more or less unphysical.
Next, we consider a wormhole solution with an exotic source confined in a region near the throat and a KS solution in the outside.
For this configuration, we try $b(r) = B \left[ 1 + \sin (\pi/16) \left( r/r_\text{th} - 1 \right) \right]$ and $\Phi(r) = (1/2) \ln (1-f(r_0)/r_0) + (1-r/r_0) (r_0f'(r_0)-f(r_0))/2(r_0-f(r_0))$ for $r\le r_0 \equiv 9r_\text{th}$ and $b(r) = 2B$ and $\Phi = (1/2) \ln (1-f(r)/r)$ for $r>r_0$, then the energy density has the following form:
\begin{equation}
\rho = \left\{ 
\begin{aligned} 
& \frac{\kappa^2\mu^2\pi B}{256r_\text{th}r^2} \cos \frac{\pi}{16}\left( \frac{r}{r_\text{th}} - 1 \right), & & \text{for } r \le r_0, \\
& 0, & & \text{for } r > r_0.
\end{aligned}
\right.
\end{equation}
The energy density is positive and confined in a sphere of the radius $r_0$, and the KS solution outside the sphere tells that $\rho=\tau=0$.
Finally, we consider a wormhole solution with an exotic source confined in a region near the throat and a Minkowski vacuum in the outside.
For this case, $b(r) = B \left[ 1 - \sin (\pi/16) \left( r/r_\text{th} - 1 \right) \right]$ for $r\le r_0$ and $b(r) = 0$ for $r>r_0$ and $\Phi(r)=0$ are considered, then the density is given by
\begin{equation}
\rho = \left\{ 
\begin{aligned} 
& -\frac{\kappa^2\mu^2\pi B}{256r_\text{th}r^2} \cos \frac{\pi}{16}\left( \frac{r}{r_\text{th}} - 1 \right), & & \text{for } r \le r_0, \\
& 0, & & \text{for } r > r_0.
\end{aligned}
\right.
\end{equation}
The confined exotic energy is negative, and the outside region is the Minkowski vacuum with $\rho=\tau=\rho_\text{HC}=\tau_\text{HC}=0$.

It is interesting to note that the exotic function of the higher curvature contribution is vanishing as $r$ grows to $r_0$ in the wormhole surrounded by KS vacuum [Fig.~\ref{fig:exotic:b}], while it becomes $-1$ as $r$ grows to the infinity in the exponential case [Fig.~\ref{fig:exotic:a}] or to $r_0$ in the wormhole surrounded by Minkowski vacuum [Fig.~\ref{fig:exotic:c}].
Since $\rho_\text{HC}>0$ for all three cases, we can rewrite the exotic function as $\zeta_\text{HC} = -1 + \tau_\text{HC}/\rho_\text{HC}$.
Then, the vanishing exotic function in Fig.~\ref{fig:exotic:b} tells us that $-\tau_\text{HC}/\rho_\text{HC} = p^r_\text{HC}/\rho_\text{HC} = -1$, which is similar to the equation of state of the vacuum energy in the cosmology, and $\zeta_\text{HC}\to-1$ in Fig.~\ref{fig:exotic:a} and \subref{fig:exotic:c} reflects $-\tau_\text{HC}/\rho_\text{HC} = p^r_\text{HC}/\rho_\text{HC} \to 0$, which becomes the equation of state of the (dark) matter.
Similarly, the exotic function of the exotic source can be written as $\zeta = 1 - \tau/\rho$ for \subref{fig:exotic:a} and \subref{fig:exotic:c} and $\zeta = -1 + \tau/\rho$ for \subref{fig:exotic:b}.
Then, the equation of state becomes $p^r/\rho \to 0$, since $\zeta\to1$ for \subref{fig:exotic:a} and \subref{fig:exotic:c}, and $p^r/\rho \to -1.5$, since $\zeta\to0.5$ for \subref{fig:exotic:b}.
Note that the equation of state for \subref{fig:exotic:b} implies that the exotic matter is a phantom energy in this case.

\section{Discussion}
\label{sec:discussion}
We have tried to build an asymptotically flat wormhole in the deformed HL gravity.
First of all, the KS wormhole can be constructed without any exotic source, but it turns out to be nontraversable since it has a horizon right at the throat.
Requiring the absence of horizon, then, the exotic source is crucial to make a traversable wormhole at least in the deformed HL theory, since it cannot be replaced by the higher curvature terms.
If DBC is relaxed, however, the higher curvature contributions might play the role of the exotic source, because it has been seen that the early acceleration of the Universe can be obtained without any inflaton fields in HL cosmology~\cite{sk}.

To make this explicit, the dark radiation and the dark scalar contributions come from the higher curvature terms of the fourth order derivative and the sixth order derivative, respectively.
Assuming that the contribution of the dark radiation is negligible, 
the contribution of the dark scalar to the energy density is crucial which is actually negative, $\rho_\text{HC}\simeq \rho_\text{ds}<0$. 
This fact can be plausible at the early stage of the Universe ~\cite{sk}. 
Now, the exotic function can be written by $\zeta_\text{HC} = 1+p^r_\text{HC}/\rho_\text{HC}$, and the equation-of-state parameter for the dark scalar turns out to be $\omega_\text{ds} = p_\text{ds}/\rho_\text{ds} = 1$. It naturally yields the positive $\zeta_\text{HC}\simeq 2>0$. What it means is that the higher curvature contribution can play the role of the exotic source when the DBC is relaxed, without resorting to the external exotic matter. Although this argument was not based on the explicit wormhole geometry but just the cosmological side, it gives some insight into the possibility to create the traversable wormhole at the early stage of the Universe in the HL gravity.  

Otherwise, the analytic continuation might be considered to resolve this issue.
Taking into account $\mu\to i\mu$ and $\zeta^2\to-i\zeta^2$~\cite{lmp}, the potential terms in the action~\eqref{action} take the opposite sign, and so do the energy density~\eqref{rho} and the tension~\eqref{tau}.
At first sight, it seems that the matter source is no more exotic and the higher curvature terms play the role of the exotic matter with this analytic continuation.
However, the relation of the speed of light~\eqref{fund:const} constrains $\omega<0$, so that $f/r$ does not vanish in the asymptotic region; instead, $f/r\approx-2r^2\omega$ and $g_{rr} = 1/(1-f/r)$ becomes large negative for $r\gg1/\sqrt\omega$.
To make $g_{rr}$ positive, Eq.~\eqref{f} can be modified to $f=-r^3\omega-\sqrt{r^3[r^3\omega^2+b(r)]}$, which approximates to $f\approx b(r)/2\omega$.
Then, it has Schwarzschild limit with $b(r\to\infty)=4\omega M<0$.
After some tedious calculations, however, the exotic function is obtained as
\begin{equation}
\zeta = - \frac{4}{r^4} \left( f + r^3\omega \right) \left[ \frac{f^2}{|b'|} \frac{d^2r}{dz^2} - \frac{r(r-f)}{|b'|} \Phi' \right].
\end{equation}
Note that $\left( f + r^3\omega \right)<0$, so the analyses on the exotic behavior are exactly same as those in Sec.~\ref{sec:exotic}, i.e. the matter source is still exotic.
Further study is needed in this issue.

Next, we have constructed three sorts of traversable wormholes: one is the wormhole with an exponentially decaying exotic source in the radial direction, another is the wormhole with an exotic source confined in the middle of the KS vacuum, and the other is the wormhole with an exotic source confined in the middle of the Minkowski vacuum.
Interestingly, in the second case, the exotic source and the higher curvature contributions, respectively, behave like the phantom energy and the dark energy near the boundary of the confined region, while in the first and third cases, both the exotic source and the higher curvature contributions behave like the (dark) matter in the boundary.

The final comment is in order.
It is interesting to note that a spherically symmetric matter distribution and the corresponding higher curvature contribution satisfy the four-dimensional covariant form of energy conservation in the deformed HL gravity, $\nabla^{(4)}_\mu T_m^{\mu\nu} = \nabla^{(4)}_\mu T_\text{HC}^{\mu\nu} = 0$, though it is obvious that the total energy satisfies the conservation of energy in the covariant form in four dimensions, $\nabla^{(4)}_\mu T^{\mu\nu} = 0$, because the left-hand sides of the equations of motion~\eqref{eom:t}--\eqref{eom:theta} are the same as the four-dimensional Einstein tensor, $\mathcal{R}^{\mu\nu} - (1/2) \mathcal{G}^{\mu\nu} \mathcal{R}$, up to the factor $4/\kappa^2$.
This behavior is not supposed to be seen in HL theory if DBC is not considered at all.
However, we should not say that it is due to DBC.
The general covariance in matter-energy conservation deserves further investigation.


\begin{acknowledgments}
This work was supported by the National Research Foundation of
Korea(NRF) grant funded by the Korea government(MEST) through the
Center for Quantum Spacetime(CQUeST) of Sogang University with grant
number 2005-0049409. W.T. Kim was also supported by the Basic Science Research Program through the  National Research Foundation of Korea(NRF) funded by the Ministry of Education, Science and Technology(2010-0008359).
\end{acknowledgments}



\begin{thebibliography}{99}

\bibitem{lifshitz}
  E.~M.~Lifshitz,
  Zh.\ Eksp.\ Teor.\ Fiz.\  {\bf 11}, 255 (1941);
%
  Zh.\ Eksp.\ Teor.\ Fiz.\  {\bf 11}, 269 (1941).

\bibitem{horava}
  P.~Ho\v{r}ava,
%
  JHEP {\bf 0903}, 020 (2009)
  [\texttt{arXiv:0812.4287 [hep-th]}];
%
  Phys.\ Rev.\  D {\bf 79}, 084008 (2009)
  [\texttt{arXiv:0901.3775 [hep-th]}];
%
  Phys.\ Rev.\ Lett.\  {\bf 102}, 161301 (2009)
  [\texttt{arXiv:0902.3657 [hep-th]}].


\bibitem{lmp}
  H.~Lu, J.~Mei and C.~N.~Pope,
  Phys.\ Rev.\ Lett.\  {\bf 103}, 091301 (2009)
  [arXiv:0904.1595 [hep-th]].

\bibitem{cco}
  R.~G.~Cai, L.~M.~Cao and N.~Ohta,
  Phys.\ Rev.\  D {\bf 80}, 024003 (2009)
  [\texttt{arXiv:0904.3670 [hep-th]}];
%
  Phys.\ Lett.\  B {\bf 679}, 504 (2009)
  [\texttt{arXiv:0905.0751 [hep-th]}].

\bibitem{cy}
  E.~O.~Colgain and H.~Yavartanoo,
  JHEP {\bf 0908}, 021 (2009)
  [\texttt{arXiv:0904.4357 [hep-th]}].

\bibitem{ks}
  A.~Kehagias and K.~Sfetsos,
  Phys.\ Lett.\  B {\bf 678}, 123 (2009)
  [\texttt{arXiv:0905.0477 [hep-th]}].

\bibitem{mipark}
  M.-I.~Park,
  JHEP {\bf 0909}, 123 (2009)
  [\texttt{arXiv:0905.4480 [hep-th]}].

\bibitem{ysmyung}
  Y.~S.~Myung,
  Phys.\ Lett.\  B {\bf 678}, 127 (2009)
  [\texttt{arXiv:0905.0957 [hep-th]}];
%
  Phys.\ Lett.\  B {\bf 684}, 158 (2010)
  [\texttt{arXiv:0908.4132 [hep-th]}];
%
  Y.~S.~Myung,
  Phys.\ Lett.\  {\bf B685}, 318-324 (2010).
  [\texttt{arXiv:0912.3305 [hep-th]}].

\bibitem{mann}
  R.~B.~Mann,
  JHEP {\bf 0906}, 075 (2009)
  [\texttt{arXiv:0905.1136 [hep-th]}].

\bibitem{gh}
  A.~Ghodsi and E.~Hatefi,
  Phys.\ Rev.\  D {\bf 81}, 044016 (2010)
  [\texttt{arXiv:0906.1237 [hep-th]}].

\bibitem{ir}
  L.~Iorio and M.~L.~Ruggiero,
  Int.\ J.\ Mod.\ Phys.\  A {\bf 25}, 5399 (2010)
  [\texttt{arXiv:0909.2562 [gr-qc]}].

\bibitem{tc}
  J.~Z.~Tang and B.~Chen,
  Phys.\ Rev.\  D {\bf 81}, 043515 (2010)
  [\texttt{arXiv:0909.4127 [hep-th]}].

\bibitem{kk:bh}
  E.~Kiritsis and G.~Kofinas,
  JHEP {\bf 1001}, 122 (2010)
  [\texttt{arXiv:0910.5487 [hep-th]}].

\bibitem{majhi}
  B.~R.~Majhi,
  Phys.\ Lett.\  B {\bf 686}, 49 (2010)
  [\texttt{arXiv:0911.3239 [hep-th]}].


\bibitem{ts}
  T.~Takahashi and J.~Soda,
  Phys.\ Rev.\ Lett.\  {\bf 102}, 231301 (2009)
  [\texttt{arXiv:0904.0554 [hep-th]}].

\bibitem{calcagni}
  G.~Calcagni,
  JHEP {\bf 0909}, 112 (2009).
  [texttt{arXiv:0904.0829 [hep-th]}].

\bibitem{kk}
  E.~Kiritsis and G.~Kofinas,
  Nucl.\ Phys.\  B {\bf 821}, 467 (2009)
  [\texttt{arXiv:0904.1334 [hep-th]}].

\bibitem{mukohyama}
  S.~Mukohyama,
  JCAP {\bf 0906}, 001 (2009)
  [\texttt{arXiv:0904.2190 [hep-th]}];
%
  S.~Mukohyama, K.~Nakayama, F.~Takahashi and S.~Yokoyama,
  Phys.\ Lett.\  B {\bf 679}, 6 (2009)
  [\texttt{arXiv:0905.0055 [hep-th]}].

\bibitem{brandenberger}
  R.~Brandenberger,
  Phys.\ Rev.\  D {\bf 80}, 043516 (2009)
  [\texttt{arXiv:0904.2835 [hep-th]}];
%
  X.~Gao, Y.~Wang, W.~Xue and R.~Brandenberger,
  JCAP {\bf 1002}, 020 (2010).
  [\texttt{arXiv:0911.3196 [hep-th]}].

\bibitem{cpt}
  B.~Chen, S.~Pi and J.~Z.~Tang,
  JCAP {\bf 0908}, 007 (2009)
  [\texttt{arXiv:0905.2300 [hep-th]}].

\bibitem{ww}
  A.~Wang and Y.~Wu,
  JCAP {\bf 0907}, 012 (2009)
  [\texttt{arXiv:0905.4117 [hep-th]}].

\bibitem{bps}
  D.~Blas, O.~Pujolas and S.~Sibiryakov,
  Phys.\ Rev.\ Lett.\  {\bf 104}, 181302 (2010)
  [\texttt{arXiv:0909.3525 [hep-th]}].

\bibitem{ls}
  G.~Leon and E.~N.~Saridakis,
  JCAP {\bf 0911}, 006 (2009)
  [\texttt{arXiv:0909.3571 [hep-th]}];
%
  S.~Carloni, E.~Elizalde and P.~J.~Silva,
  Class.\ Quant.\ Grav.\  {\bf 27}, 045004 (2010)
  [\texttt{arXiv:0909.2219 [hep-th]}].

\bibitem{sj}
  M.~R.~Setare and M.~Jamil,
  JCAP {\bf 1002}, 010 (2010)
  [\texttt{arXiv:1001.1251 [hep-th]}];
%
  M.~Jamil, E.~N.~Saridakis and M.~R.~Setare,
  \textit{The generalized second law of thermodynamics in Horava-Lifshitz
  cosmology},
  \texttt{arXiv:1003.0876 [hep-th]}.

\bibitem{sk}
  E.~J.~Son, W.~Kim,
  JCAP {\bf 1006}, 025 (2010).
  [\texttt{arXiv:1003.3055 [hep-th]}].

\bibitem{sk2}
  E.~J.~Son, W.~Kim,
  Mod.\ Phys.\ Lett.\ A {\bf 26}, 719-725 (2011)
  \texttt{arXiv:1007.5371 [gr-qc]}.


\bibitem{mt}
  M.~S.~Morris and K.~S.~Thorne,
  Am.\ J.\ Phys.\  {\bf 56}, 395 (1988).
%


\bibitem{adm}
  R.~L.~Arnowitt, S.~Deser and C.~W.~Misner,
  Phys.\ Rev.\  {\bf 117}, 1595 (1960);
%
  ``The dynamics of general relativity''
  in \textit{Gravitation: an introduction to current research}, ed. L.~Witten pp 227-265 (New York: Wiley 1962)
  [\texttt{gr-qc/0405109}].


\bibitem{djt}
  S.~Deser, R.~Jackiw and S.~Templeton,
  Annals Phys.\  {\bf 140}, 372 (1982)
  [Erratum-ibid.\  {\bf 185}, 406 (1988); ibid.\ {\bf 281}, 409-449 (2000)];
%
  Phys.\ Rev.\ Lett.\  {\bf 48}, 975 (1982).


%

\end{thebibliography}
\end{document}